\documentclass[twocolumn,showpacs,preprintnumbers,amsmath,amssymb]{revtex4}
\usepackage{amsfonts}

\usepackage{graphicx}
\usepackage{bm}

\newcommand{\beq}{\begin{equation}}
\newcommand{\eeq}{\end{equation}}
\newcommand{\beqa}{\begin{eqnarray}}
\newcommand{\eeqa}{\end{eqnarray}}
\newcommand{\beqar}{\begin{eqnarray*}}
\newcommand{\eeqar}{\end{eqnarray*}}

\newcounter{saveeqn}

\begin{document}
\title{The security of Ping-Pong protocol}
\author{Jian-Chuan Tan}\email{tanjc@mail.ustc.edu.cn}
\author{An Min Wang}
\affiliation{Quantum Theory Group, Department of Modern Physics\\
University of Science and Technology of China, Hefei, 230026,
P.R.China}

\begin{abstract}
Ping-Pong protocol is a type of quantum key distribution which makes
use of two entangled photons in the EPR state. Its security is based
on the randomization of the operations that Alice performs on the
travel photon (qubit), and on the anti-correlation between the two
photons in the EPR state. In this paper, we study the security of
this protocol against some known quantum attacks, and present a
scheme that may enhance its security to some degree.
\end{abstract}


\maketitle

\section{INTRODUCTION}

In the early of 1990s, A. K. Ekert proposed a conception of
realizing quantum cryptography based on Bell's theorem \cite{Ekert},
and C. H. Bennett and S. J. Wiesner brought forward a scheme for
communicating via one- and two-particle operators on EPR states
\cite{Bennett1}. Since then, how to use EPR pairs to distribute a
key has become a significant field of quantum key distribution (QKD)
and has drawn physicists's attention. In 2002, Kim Bostr\"{o}m and
Timo Felbinger proposed a novel QKD protocol called `Ping-Pong'
protocol \cite{Bostrom}, a number of works have been done in this
aspect of QKD by far, some of them suggested improving security
level of it, while others aimed at proposing eavesdropping schemes
to attack it. In this letter, we discuss its robustness to  some
known quantum attacks in order to render a general description of
this protocol.

For the purpose, let us recapitulate the `Ping-Pong' protocol: Bob
prepares two photons in an entangled state $\left| {\psi^+}
\right\rangle = \frac{1}{\sqrt 2 }(\left| 0 \right\rangle \left| 1
\right\rangle + \left| 1 \right\rangle \left| 0 \right\rangle )$. He
keeps one of them (home photon), and sends the other (travel
photon), to Alice through a quantum channel. After receiving the
travel photon, Alice randomly switches between a control mode and a
message mode. In the control mode, Alice measures the travel photon
with basis $B _z=\{ \left| 0 \right\rangle , \left| 1 \right\rangle
\} $ and then announces her measurement result through a classical
public channel. After receiving the public announcement from Alice,
Bob also switches to the control mode and measures the home photon
with the same basis. In the absence of an eavesdropper, Eve, both
the results should be anti-correlated, otherwise, it is an evidence
that Eve is in line, the QKD process should be stopped. In the
message mode, Alice performs a unitary operation $Z^j$ to encode her
message $j \in \{0,1\}$ on the travel photon, where $Z^j = \left| 0
\right\rangle \left\langle 0 \right| +(-)^j \left| 1 \right\rangle
\left\langle 1 \right|$. Then Alice sends the travel photon to Bob.
Bob performs a measurement with a Bell basis to draw the information
Alice encoded. If the measurement result is $\left| {\psi^+}
\right\rangle$, Bob knows that $j=0$. Likewise, if the result is
$\left| {\psi^-} \right\rangle $, Bob knows that $j=1$. Repeating
the process above could transmit the classic bits that Alice would
like to share with Bob, so the QKD is done. The security of
Ping-Pong protocol is based on the randomization of the operations
that Alice performs on the travel photon (qubit), and on the
anti-correlation between the two photons in the EPR state.

Compared to BB84 \cite{Bennett2} and B92 \cite{Bennett3}, Ping-Pong
protocol possesses a remarkable advantage: in the QKD process, it is
unnecessary for Alice and Bob to discard some (may be a considerable
amount of) unsuitable bits, so the efficiency of Ping-Pong protocol
was ever thought of to be higher than BB84 and B92 by some
researchers. However, `how safe is it' is still a problem that needs
to be solved. In the next three sections, I'll discuss this
problem.\\

\section{TO OPAQUE EAVESDROPPING}

The opaque eavesdropping is the simplest attack, which is also
called `intercept-resend attack'. In this eavesdropping, Eve
intercepts the quantum carrier on its way from Alice to Bob and/or
from Bob to Alice and performs a measurement to get information
about what state is sent and which operation Alice performs to the
travel photon. Fig.1 demonstrates Eve's eavesdropping process.
\begin{figure}[htbp] \centerline{\includegraphics[width=3.00in,
height=0.80in]{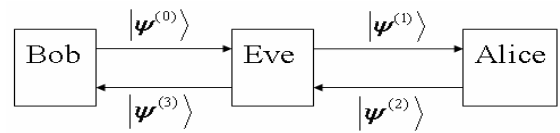}}  \caption{The process of Eve's
eavesdropping.} \label{fig1}
\end{figure}

Ping-Pong protocol itself plus this opaque eavesdropping could be
described as follows:\\
(1) At first, Bob prepares a pair of photons in $\left| {\psi ^ +
} \right\rangle = \frac{1}{\sqrt 2}(\left| 0 \right\rangle \left|
1 \right\rangle + \left| 1 \right\rangle \left| 0 \right\rangle
)$. Assuming that Eve is absent, then after Alice's opration on
the travel photon($Z^0$ with the probability $p_o$, and $Z^1$ with
$p_1$), the state of the pair becomes
\begin{equation} \left|\psi ''
\right\rangle = {\sqrt {p_{_0}}} \left| {\psi ^ + } \right\rangle +
{\sqrt {p_{_1}}} \left| {\psi ^ - } \right\rangle = \frac{1} {\sqrt
2} \left( {{\begin{array}{*{20}c}
\hfill 0 \hfill \\
{\sqrt {p_0}} + {\sqrt {p_1}} \hfill \\
{\sqrt {p_0}} - {\sqrt {p_1}} \hfill \\
\hfill 0 \hfill \\
\end{array} }} \right).
\end{equation}
(2) When Eve is in line, she captures the travel photon from Bob
to Alice, and performs a projective measurement on it. Assuming Eve
gets $\left| 0 \right\rangle$, the home qubit at Bob's hand becomes
$\left| 1 \right\rangle$. Then Eve prepares another $\left| 0
\right\rangle$ and sends it to Alice. Alice operates $Z^0$ (with
$p_0$) or $Z^1$ (with $p_1$) to the qubit. However, neither $Z^0$
nor $Z^1$ could change $\left| 0 \right\rangle$. After that, Alice
returns the qubit, and Eve captures it. But Eve could not get any
information by performing measurement on it because it remains
unchanged. At last, Bob receives the travel qubit and the final
state of the pair reads $\left| 1 \right\rangle _ h \left| 0
\right\rangle _ t$. In this case, the probability that Bob gets the
wrong information, i.e. QBER, could be expressed as
$q_{_0}=1-|\langle \psi ''|1 0 \rangle|^2=\frac{1}{2}-{\sqrt {p_{_0}
p_{_1}}}=\frac{1}{2}+{\sqrt {p_{_0} (1-p_{_0})}}$. \\
(3) Similarly, The case that Eve gets $\left| 1 \right\rangle$ when
she performs measurement on the qubit from Bob to Eve can be
analyzed analogously as above. But the QBER is
$q_{_1}=\frac{1}{2}-{\sqrt {p_{_0} (1-p_{_0})}}$.

Since the probability that Eve gets $\left| 0 \right\rangle$ or
$\left| 1 \right\rangle$ is $50\%$ respectively, the statistical
QBER is $q = \frac{q_{_0} + q_{_1}} {2} = \frac{1}{2}$. In this
case, the capacity of the channel, i.e. the maximal information
between Alice and Bob, can be calculated to be \beq I_{AB} = 1 + q
\log_2 q + (1-q) \log_2 (1-q) = 0.\eeq

From the analysis above, one can come to a conclusion that opaque
eavesdropping is unskilled to Ping-Pong protocol, because it could
make neither Eve nor Bob obtain any information, and in addition, it
may cause a QBER up to $50\%$. As a result, a wise eavesdropper
would not use it, so Ping-Pong protocol is robust to opaque attack.

\section{TO TRANSLUCENT EAVESDROPPING}

We will still follow the process in Sec.\textbf{II} (see Fig.1). In
the following analysis, we make an assumption that measurements do
not make photons disappear, although with current technology, a
photon disappears after it is measured. In fact, if the photon
disappears, the analysis in this section would degenerate to that in
Sec.\textbf{II}.

(1) Bob prepares a pair of qubits in $\left| \psi ^ {(0)}
\right\rangle = \left| \psi ^+ \right\rangle = \frac{1}{\sqrt 2} (
\left| 0 \right\rangle _ h \left| 1 \right\rangle _t + \left| 1
\right\rangle _ h \left| 0 \right\rangle _ t )$, and sends one of
them to Alice.\\

(2) Eve captures the travel qubit, and makes it interact with an
ancilla $\left| \chi \right\rangle$, obtaining
\begin{widetext}
\begin{equation}
\begin{array}{lll}
\left| \psi^{(1)} \right\rangle& = &\frac{1}{\sqrt 2} [ \left| 0
\right\rangle _ h ( {\sqrt F} \left| 1 \right\rangle _ t \left| \chi
_ 1 \right\rangle + {\sqrt D} \left| 0 \right\rangle _ t \left| \chi
_0 \right\rangle ) + \left| 1 \right\rangle _ h ( {\sqrt F} \left| 0
\right\rangle _ t \left| \chi _ 0 \right\rangle + {\sqrt D} \left| 1
\right\rangle _ t \left| \chi _1 \right\rangle )] \\
&= &( \sqrt {\frac{D}{2}} \left| 0 \right\rangle _ h + \sqrt
{\frac{F}{2}} \left| 1 \right\rangle _ h ) \left| 0 \right\rangle _
t \left| \chi _ 0 \right\rangle + ( \sqrt {\frac{F}{2}} \left| 0
\right\rangle _ h + \sqrt {\frac{D}{2}} \left| 1 \right\rangle _ h )
\left| 1 \right\rangle _ t \left| \chi _ 1 \right\rangle)
\end{array},
\end{equation}
\end{widetext}
\begin{widetext}
in which $D$ is the probability of error, and $F+D=1$. Thus the
reduced density matrix of the ancilla and travel qubit reads
\begin{equation}
\rho _{at}^{(1)} = tr^h \left| \psi^{(1)} \right\rangle
\left\langle \psi^{(1)} \right| = \frac{1}{2} \left(
{{\begin{array}{*{40}c}
1 & 0 & 0 & 2\sqrt{D(1-D)} \\
0 & 0 & 0 & 0\\
0 & 0 & 0 & 0\\
2\sqrt{D(1-D)} & 0 & 0 & 1 \\
\end{array} }} \right)
\end{equation}
\end{widetext}

(3) Eve continues to pass the travel qubit to Alice. If Alice
chooses 'control mode', she would detect out Eve with a probability
$D$. Else, if Alice chooses 'message mode', performing $Z^0$ with
$p_{_0}$ and $Z^1$ with $p_{_1} = 1 - p_{_0}$, the state of the
ancilla and travel qubit would become
\begin{widetext}
\begin{equation}
\begin{array}{lll}
\left| \psi^{(2)} \right\rangle &= &\frac{\sqrt{D} \left| 0
\right\rangle _h + \sqrt{F} \left| 1 \right\rangle _h}{\sqrt 2}
(\sqrt{p_{_0}} + \sqrt{p_{_1}}) \left| 0 \right\rangle _t \left|
\chi _0 \right\rangle + \frac{\sqrt{F} \left| 0 \right\rangle _h +
\sqrt{D} \left| 1 \right\rangle _h}{\sqrt 2} (\sqrt{p_{_0}} -
\sqrt{p_{_1}}) \left|
1 \right\rangle _t \left| \chi _{_1} \right\rangle\\
&= &\left| 0 \right\rangle _h [ \sqrt{\frac{D}{2}} ( \sqrt{p_{_0}} +
\sqrt{p_{_1}} ) \left| 0 \right\rangle _t \left| \chi _{_0}
\right\rangle + \sqrt{\frac{F}{2}} ( \sqrt{p_{_0}} - \sqrt{p_{_1}} )
\left| 1 \right\rangle _t \left| \chi _{_1} \right\rangle ]\\
&+ & \left| 1 \right\rangle _h [ \sqrt{\frac{F}{2}} ( \sqrt{p_{_0}}
+ \sqrt{p_{_1}} ) \left| 0 \right\rangle _t \left| \chi _{_0}
\right\rangle + \sqrt{\frac{D}{2}} ( \sqrt{p_{_0}} - \sqrt{p_{_1}} )
\left| 1 \right\rangle _t \left| \chi _{_1} \right\rangle
]\end{array}.
\end{equation}
\end{widetext}
The density matrix is
\begin{widetext}
\begin{equation} \rho _{at}^{(2)} = \left(
{{\begin{array}{*{40}c}
\frac{1}{2} + \sqrt{ p_{_0} p_{_1}} & 0 & 0 & ( p_{_0} - p_{_1} ) \sqrt{D(1-D)} \\
0 & 0 & 0 & 0\\
0 & 0 & 0 & 0\\
( p_{_0} - p_{_1} ) \sqrt{D(1-D)} & 0 & 0 & \frac{1}{2} - \sqrt{ p_{_0} p_{_1}} \\
\end{array} }} \right),\end{equation}
\end{widetext}
whose eigenvalues are as follows
\begin{equation} \begin{array}{lll}
\lambda _1 = \lambda _2 = 0,\\
\lambda _3 = \frac{1}{2} + \sqrt { p_{_0} p_{_1} + ( p_{_0} -
p_{_1} )^2 D \left(1-D\right)}, \\
\lambda _4 = \frac{1}{2} - \sqrt { p_{_0} p_{_1} + ( p_{_0} - p_{_1}
)^2 D \left(1-D\right)}.\end{array}\end{equation} Thus the maximum
information Eve could get can be calculated as \cite{Bostrom}
\begin{equation} I_{AE} = - \sum_{i=1}^4 \lambda_i \log_2
\lambda_i = - \lambda_3 \log_2 \lambda_3 - \lambda_4 \log_2
\lambda_4.\end{equation}\\

In fact, the density matrix of the whole system (ancilla, home and
travel qubit) reads
\begin{widetext}
\begin{equation} \rho^{(2)} = \frac{1}{2} \left(
{{\begin{array}{*{40}c}
D & 0 & 0 & (p_{_0} - p_{_1})\sqrt{D(1-D)} & \sqrt{D(1-D)} & 0 & 0 & (p_{_0} - p_{_1})D \\
0 & 0 & 0 & 0 & 0 & 0 & 0 & 0\\
0 & 0 & 0 & 0 & 0 & 0 & 0 & 0\\
(p_{_0} - p_{_1})\sqrt{D(1-D)} & 0 & 0& 1-D & (p_{_0} - p_{_1})(1-D) & 0 & 0 & \sqrt{D(1-D)} \\
\sqrt{D(1-D)} & 0 & 0 & (p_{_0} - p_{_1})(1-D) & 1-D & 0 & 0 & (p_{_0} - p_{_1})\sqrt{D(1-D)} \\
0 & 0 & 0 & 0 & 0 & 0 & 0 & 0\\
0 & 0 & 0 & 0 & 0 & 0 & 0 & 0\\
(p_{_0} - p_{_1})D & 0 & 0& \sqrt{D(1-D)} & (p_{_0} - p_{_1})\sqrt{D(1-D)} & 0 & 0 & D \\
\end{array} }} \right).\end{equation}
\end{widetext}

(4) The travel qubit is sent back by Alice and captured again by
Eve. Now, Eve could perform a measurement on the two qubits: ancilla
and the travel one. In this case, this qubit pair is in the EPR
state in the subspace of the two qubits, thus a Bell measurement
could help Eve get the information about which operation Alice has
performed on the travel one. Eve makes use of two Bell basis-vectors
for the measurement:
\begin{equation} \begin{array}{lll}
\left| \phi_{at}^I \right\rangle = \frac{1}{\sqrt 2} ( \left| 0
\right\rangle _t \left| \chi _{_0} \right\rangle + \left| 1
\right\rangle _t \left| \chi _{_1} \right\rangle )\\
\left| \phi_{at}^Z \right\rangle = \frac{1}{\sqrt 2} ( \left| 0
\right\rangle _t \left| \chi _{_0} \right\rangle - \left| 1
\right\rangle _t \left| \chi _{_1} \right\rangle ). \end{array}
\end{equation} So the probability of Alice's operation that Eve
obtains could be calculated as
\begin{equation}\begin{array}{lll}
p(I) = \left\langle \phi_{at}^I \right| \rho_{at}^{(2)} \left|
\phi_{at}^I \right\rangle = \frac{1}{2} + ( p_{_0} - p_{_1} ) \sqrt{
D(1-D)}\\
p(Z) = \left\langle \phi_{at}^Z \right| \rho_{at}^{(2)} \left|
\phi_{at}^Z \right\rangle = \frac{1}{2} - ( p_{_0} - p_{_1} ) \sqrt{
D(1-D)}.\end{array}
\end{equation}
After Eve's Bell measurement, the subsystem of home and travel
qubits becomes
\begin{widetext}
\begin{equation}
\rho_{ht}^{(3)} = \left( {{\begin{array}{*{30}c}
\frac{1}{4} + \frac{1}{2} \sqrt{p_{_0} ( 1 - p_{_0})} (2D-1) & 0 & \frac{1}{2}\sqrt{D(1-D)} & 0 \\
0 & \frac{1}{4} + \frac{1}{2} \sqrt{p_{_0} ( 1 - p_{_0})} (2D-1) & 0 & \frac{1}{2}\sqrt{D(1-D)}\\
\frac{1}{2}\sqrt{D(1-D)} & 0 & \frac{1}{4} - \frac{1}{2} \sqrt{p_{_0} ( 1 - p_{_0})} (2D-1) & 0\\
0 & \frac{1}{2}\sqrt{D(1-D)} & 0 & \frac{1}{4} - \frac{1}{2} \sqrt{p_{_0} ( 1 - p_{_0})} (2D-1) \\
\end{array} }} \right).
\end{equation}
\end{widetext} Note that $D+F=1$ and $p_{_0} + p_{_1} =1$.

(5) Eve sends the travel qubit back to Bob. Statistically, Bob uses
performs measurements on both home and travel qubits, and the QBER
of Bob's measurements is
\begin{equation}
q = 1 - \left\langle \psi '' \right| \rho_{ht}^{(3)} \left| \psi ''
\right\rangle = \frac{3}{4} - p_{_0} (1 - p_{_0}) (2D-1).
\end{equation}
Thus, the maximum mutual information between Alice and Bob (i.e. the
capacity of this quantum channel) is \cite{Ekert2}
\begin{equation}
I_{AB} = 1 + q \log_2 q + (1-q) \log_2 (1-q).
\end{equation}
We plot $I_{AE}$ and $I_{AB}$ in the figure below:
\begin{figure}[htbp] \centerline{\includegraphics[width=3.00in,
height=2.50in]{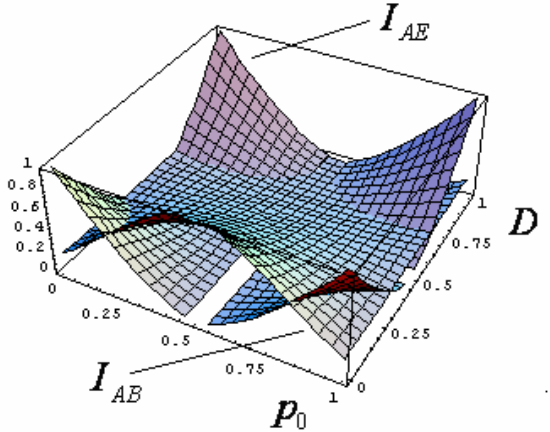}}  \caption{The comparison of $I_{AB}$ and
$I_{AE}$.}
\label{fig2}\end{figure}\\

From the Fig.2, we can come to a conclusion that when Alice conducts
an equiprobable coding, that is to say, $p_{_0}=p_{_1}=\frac{1}{2}$,
which is also the security requirement of classical cryptography
\cite{Stallings}, Bob could always gets more information than Eve,
especially in the case of $D=0$ (Eve's optimal eavesdropping), Alice
and Bob can share the maximum information ($I_{AB} = 1$), and Eve
gets the minimum information ($I_{AE} = 0$). Thus, in short, as long
as Alice codes the travel qubits equiprobably, the Ping-Pong
protocol is secure.

\section{TO W\'{o}jcik's ATTACK}

In 2003, Antoni W\'{o}jcik proposed a novel eavesdropping scheme to
attack Ping-Pong protocol \cite{Wojcik}, claiming that if the
quantum channel transmission efficiency $\eta$ is no more than
$60\%$, Eve could get more information than $I_{AB}$ without being
detected. By far, there is no effective preventing method against
this attack. We propose a so-called Disguising Photon Detecting
(DPD) method to implement this task, because in our scheme, we use
some single photons in the state $\left| + \right\rangle =
\frac{1}{\sqrt 2}(\left| 0 \right\rangle + \left| 1 \right\rangle )$
disguising the travel photons in the original Ping-Pong protocol.
The disguising photon can be called a `false photon', and
correspondingly, the travel photon entangled with another one in the
state $\left| {\psi ^ + } \right\rangle = \frac{1}{\sqrt 2}(\left| 0
\right\rangle \left| 1 \right\rangle + \left| 1 \right\rangle \left|
0 \right\rangle )$ in the original Ping-Pong protocol \cite{Bostrom}
is called a `true photon'. In the DPD method, the original Ping-Pong
protocol has to be modified: Bob randomly sends Alice a travel
photon that is a true or false photon. After receiving the travel
photon, Alice switches between control mode and message mode, and
then goes ahead just as in the original Ping-Pong protocol. But
after Bob receiving the traveling-back photon, what he should do is
somewhat different from that in the original Ping-Pong protocol. If
he sends a true photon, he then take the same action on the
traveling-back photon just as what he should do in the original
Ping-Pong protocol; else, if he sends a false photon and Alice
chooses the message mode, after receiving the traveling-back photon,
he asks Alice which operation she performed on the photon, $Z^0$ or
$Z^1$, if Alice performed $Z^0$, Bob does nothing to the
traveling-back photon and discard it, while if Alice performed
$Z^1$, Bob performs a projective measurement on the traveling-back
photon with the projector $P_+ = \left| + \right\rangle \left\langle
+ \right| = \frac{1}{2}(\left| 0 \right\rangle + \left| 1
\right\rangle )(\left\langle 0 \right| + \left\langle 1 \right|)$,
which could be done by using some optical devices
\cite{Nielson,Huang,Zhang'}; else, if Bob sends a false photon and
Alice chooses the control mode, Bob tells Alice to discard this bit
in the authentication step after sending all the photons.

Theoretically, if Eve is absent, the false photon after being
performed $Z^1$ must be in the state $\left| - \right\rangle =
\frac{1}{\sqrt 2}(\left| 0 \right\rangle - \left| 1
\right\rangle$, so the outcome of the measurement must be zero,
because $\left| - \right\rangle $ is orthogonal to $\left| +
\right\rangle $. But if the outcome is not zero, it could be an
evidence that Eve is eavesdropping the communication between Alice
and Bob, thus the QKD process must be stopped.

To demonstrate this method is feasible, let us now analyze the
states that Bob sends and receives. The initial state that Bob
sends is $\left| + \right\rangle = \frac{1}{\sqrt 2 }(\left| 0
\right\rangle + \left| 1 \right\rangle )$, whose density matrix is
\begin{equation}
 \rho _{_{\scriptstyle t}} = \left| + \right\rangle
\left\langle + \right| = \frac{1}{2}(\left| 0 \right\rangle
\left\langle 0 \right| + \left| 0 \right\rangle \left\langle 1
\right| + \left| 1 \right\rangle \left\langle 0 \right| + \left| 1
\right\rangle \left\langle 1 \right|) = \frac{1}{2}\left(
{{\begin{array}{*{20}c}
 1 \hfill & 1 \hfill \\
 1 \hfill & 1 \hfill \\
\end{array} }} \right).
\end{equation}

There are two situations that should be considered:

(1) When W\'{o}jcik's Eve is in line. After Eve's $B\mbox{-}A$
attack, the state becomes $\left|{B\mbox{-}A} \right\rangle =
Q_{txy} \left| + \right\rangle _t \left| {vac} \right\rangle _x
\left| 0 \right\rangle _y = \frac{1}{2}(\left| 0 \right\rangle _t
\left| 0 \right\rangle _x \left| {vac} \right\rangle _y + \left|
{vac} \right\rangle _t \left| 0 \right\rangle _x \left| 1
\right\rangle _y ) + \frac{1}{2}(\left| {vac} \right\rangle _t
\left| 1 \right\rangle _x \left| 0 \right\rangle _y + \left| 1
\right\rangle _t \left| 1 \right\rangle _x \left| {vac}
\right\rangle _y )$. If Alice performs $Z^0$ (that is, an identical
operation $I$), the state maintains in
$\left|B\mbox{-}A\right\rangle $, then Alice sends it back and Eve
commits the $A\mbox{-}B$ attack $Q^{-1}_{txy}$ (according to
transformations (3) in W\'{o}jcik's paper \cite{Wojcik}):
$\left|{A\mbox{-}B} \right\rangle _I = Q^{-1}_{txy} I \left|
B\mbox{-}A \right\rangle= \frac{1}{\sqrt 2}(\left| 0 \right\rangle
_t \left| {vac} \right\rangle _x \left| 0 \right\rangle _y + \left|
1 \right\rangle _t \left| {vac} \right\rangle _x \left| 0
\right\rangle _y = \left| + \right\rangle _t \left| {vac}
\right\rangle _x \left| 0 \right\rangle _y)$, where the subscript
$I$ indicates that Alice performs $I = Z^0$. So the travel photon
Bob receives would be still in the state $\left| + \right\rangle $.
Else, if Alice performs $Z^1$ (that is, the Pauli $\sigma_z$
operation), the state becomes $ \sigma_z \left| B\mbox{-}A
\right\rangle = \frac{1}{2}(\left| 0 \right\rangle _t \left| 0
\right\rangle _x \left| {vac} \right\rangle _y + \left| {vac}
\right\rangle _t \left| 0 \right\rangle _x \left| 1 \right\rangle
_y) + \frac{1}{2}(\left| {vac} \right\rangle _t \left| 1
\right\rangle _x \left| 0 \right\rangle _y - \left| 1 \right\rangle
_t \left| 1 \right\rangle _x \left| {vac} \right\rangle _y)$. Then
Alice sends the photon back and Eve commits the $A\mbox{-}B$ attack
$Q^{-1}_{txy}$: $\left|{A\mbox{-}B} \right\rangle _Z = Q^{-1}_{txy}
\sigma_z \left| B\mbox{-}A \right\rangle = \frac{1}{\sqrt 2}\left| 0
\right\rangle _t \left| {vac} \right\rangle _x \left| 0
\right\rangle _y + \frac{1}{\sqrt 2}\left| 1 \right\rangle _t \left|
{vac} \right\rangle _x \left| 1 \right\rangle _y$, where the
subscript $Z$ indicates that Alice performs $Z = Z^1$. So the
density matrix of the false photon is \beq \rho _{_{\scriptstyle
Zt}} = Tr_{x,y} \left| A\mbox{-}B \right\rangle _Z { } _Z
\left\langle A\mbox{-}B \right| = \frac{1}{2}\left| 0 \right\rangle
\left\langle 0 \right| + \frac{1}{2}\left| 1 \right\rangle
\left\langle 1 \right|\eeq where the subscript `$t$' denotes the
travel photon. This means that the false photon Bob receives is in
either $\left| 0 \right\rangle$ or $\left| 1 \right\rangle$ with the
probability of 1/2 respectively.

(2) When W\'{o}jcik's Eve is absent. If Alice performs $Z^0$ on
the false photon and sends it back, Bob would receives the photon
in $\left| + \right\rangle$. Else, if Alice performs $Z^1$, the
photon Bob receives would be in $Z^1 \left| + \right\rangle =
\left| - \right\rangle$.

It could be concluded from the analysis above that if Alice performs
$Z^0$ on the false photon, no matter W\'{o}jcik's Eve is in line or
not, Bob would receive the photon in $\left| + \right\rangle$, which
is not able to be used to detect Eve. But if Alice performs $Z^1$,
the case is different: when Eve is in line, Bob would receives the
photon in either $\left| 0 \right\rangle$ or $\left| 1
\right\rangle$; when Eve is absent, Bob would receives the photon in
$\left| - \right\rangle$. With this difference, to detect
W\'{o}jcik's Eve is possible, and we propose
a projector $P_+$ could fulfill this task.\\

\section{DISCUSSION AND COMMENT}

In 2003, Qing-yu Cai published his comment claiming that the
Ping-Pong protocol can be attacked without eavesdropping \cite{Cai}.
In the comment, Cai proposed that Eve could attack the communication
between Alice and Bob with the following method: `{\em In every
message mode, Eve captures the travel back qubit Alice sent to Bob
and perform a measurement in the basis $B_z $ and forwards to Bob
this qubit. Alice and Bob have zero probability to find Eve's
attack. Then Bob lets this communication continue. But every one of
Bob's measurement results is meaningless since the two qubits become
independent of each other after Eve's attack measurement\ldots .
When the communication is terminated, Bob has learned nothing but a
sequence of nonsense random bits.}' However, we think this attack
would not work as well as claimed for at least two reasons: (1),
when Eve performs a measurement on the qubit travelling from Alice
to Bob, the entanglement between the home qubit and the travel qubit
is destroyed, it is no longer $\left| {\psi ^ + } \right\rangle =
\frac{1}{\sqrt 2 }(\left| 0 \right\rangle \left| 1 \right\rangle +
\left| 1 \right\rangle \left| 0 \right\rangle )$ or $\left| {\psi ^
- } \right\rangle = \frac{1}{\sqrt 2 }(\left| 0 \right\rangle \left|
1 \right\rangle - \left| 1 \right\rangle \left| 0 \right\rangle )$,
but simply $\left| {01} \right\rangle $ or $\left| {10}
\right\rangle $, which could be detected by a Bell measurement, and
Eve could not gain any useful information about what operation Alice
performs on the travel qubit; (2), after terminating the QKD
process, Alice and Bob would pick out a part of the key established
in the process to make a classic authentication, if the attack makes
Bob's measurement results meaningless, it would be found that Eve is
in line in the classic authentication. As a result, Cai's claim that
the Ping-Pong protocol can be attacked without eavesdropping is open to doubt.\\

\section{CONCLUSION}

\begin{figure*}[htbp] \centerline{\includegraphics[width=7.00in,
height=5.50in]{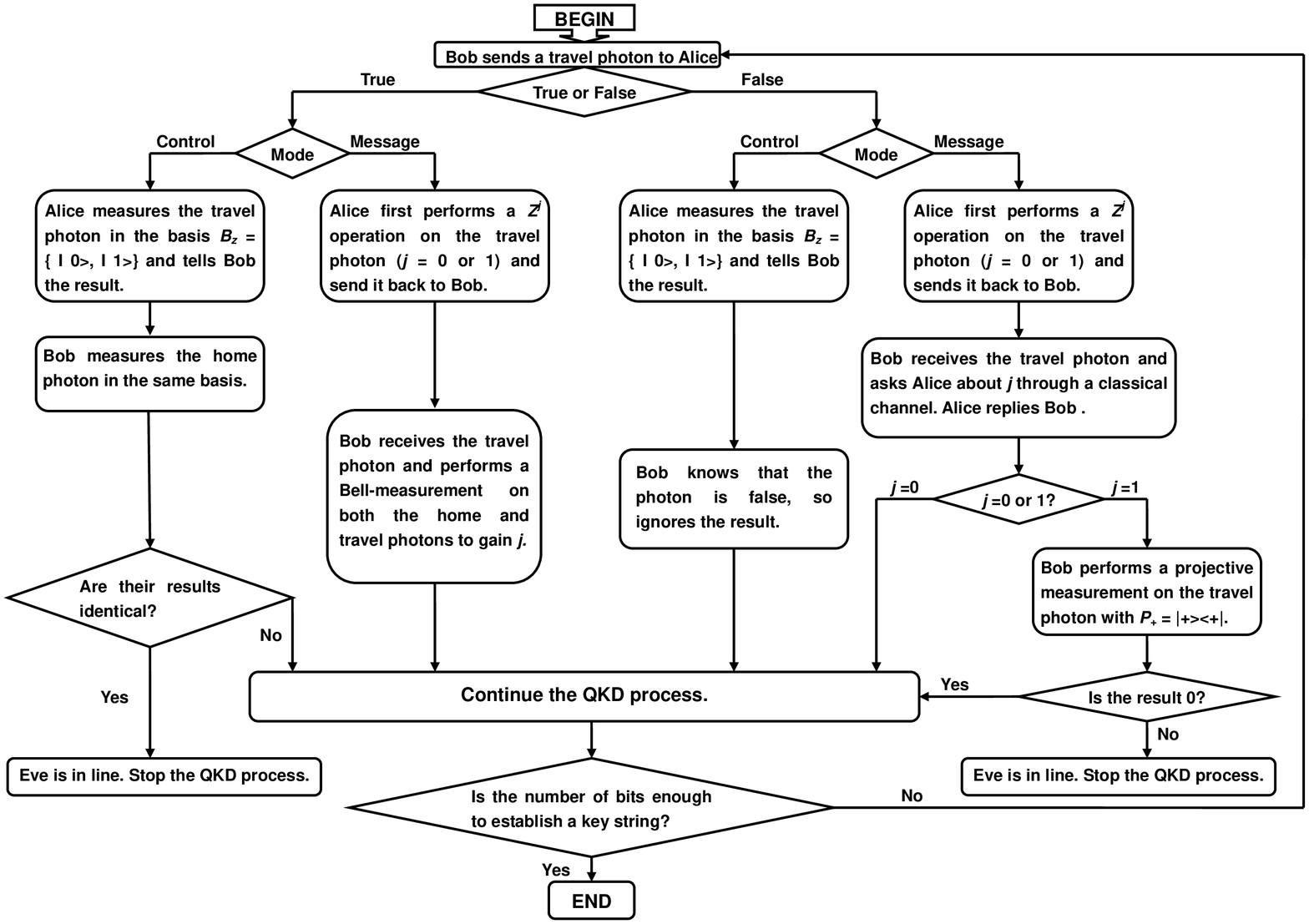}}  \caption{The flow chart of the modified
Ping-Pong protocol.} \label{fig3}
\end{figure*}

In this letter, we analyze the robustness of Ping-Pong protocol to
some known quantum attacks, from the analysis, we can come to the
conclusion that, to opaque and translucent attacks, Ping-Pong
protocol is robust and secure, and to W\'{o}jcik's attack, as long
as Bob sends sufficient disguising photons, he could make this
attack useless. In summary, the Ping-Pong protocol is secure as long
as it is modified to use the DPD method. We call the Ping-Pong
protocol associated with the DPD method a modified Ping-Pong
protocol, of which process may not be depicted clearly in words, so
it would be necessary and beneficial to describe it in a chart.
Thus, we draw a flow chart to make the modified Ping-Ping protocol
more clear to be understood. See Fig.\ref{fig3}:\\

{\em Acknowledgement: } We are grateful to all the collaborators of
our quantum theory group in the institute for theoretical physics of
my university. This work was supported by the National Natural
Science Foundation of China under Grant No. 60573008.

\end{document}